\documentclass[twocolumn,aps,prl,floatfix,showpacs]{revtex4}
\usepackage{graphicx}

\begin{document}

\title{
Origin of $sp$-electron magnetism in Graphitic Carbon Nitride
}

\author{Wei Xu, Jin Shang, Jie-Xiang Yu},
\author{J. G. Che}\altaffiliation{Corresponding author.
E-mail: jgche@fudan.edu.cn.}

\affiliation{Surface Physics Laboratory (National Key Laboratory),
Key Laboratory of Computational Physical Sciences (MOE),Fudan University,
Shanghai 200433, People's Republic of China
and Collaborative Innovation Center of Advanced Microstructures, Nanjing 210093, China
}


\begin{abstract}
Based on first principles calculations, this study reveals that magnetism in otherwise non-magnetic
materials can originate from the partial occupation of antibonding states. Since the antibonding
wavefunctions are spatially antisymmetric, the spin wavefunctions should be
symmteric according to the exchange antisymmetric principle of quantum mechanics. We
demonstrate that this phenomenon can be observed in a graphitic carbon nitride material,
$g$-C$_4$N$_3$, which can be experimentally synthesized and seen as a honeycomb structure
with a vacancy. Three dangling bonds of N atoms pointing to the vacancy site interact with each
other to form one bonding and two antibonding states. As the two antibonding states are near the
Fermi level, and electrons should partially occupy the antibonding states in spin polarization, this
leads to 1~$\mu_B$ magnetic moment. 
\end{abstract}

\maketitle

Theoretical predictions of the magnetism of otherwise non-magnetic materials (induced by
$sp$-electrons) have been reported for four decades, and some of these predictions have been
experimentally confirmed, see e.g. Ref.~\cite{Mak06,Yaz10,Vol10}. Research on the magnetism has led
to a large of publications and patents~\cite{Mak06}, since the advantages of the magnetism of
$sp$-electrons compared with the magnetism by $d$- or $f$-electrons (of transition metal [TE] or
rare-earth metal [REM]) are vital to developing spin-based 
electronics~\cite{Yaz10,Vol10,Pes12,Esq13,Khe14,Han14,Hol16,Nig17}. 
However,
although a number of impressive findings have been accumulated that have been of significance
within both the science and technology domains, there remains a distinct lack of understanding of
the origin of magnetism in such materials and this subject continues to puzzle
researchers~\cite{Mak06,Yaz10,Ohl07,Vol10,Elf02,Wan08,Baz11,Du12,Pal12,Wan09,Nai12,Kum14}. 
40 years ago experiments that reported magnetism in materials without TM or REM 
were often criticized on the basis that their
studies were not conducted carefully enough, and the findings may have been the result of
magnetic impurities~\cite{Mak06}.
Although more and more experimental evidence came to light that the spontaneous
magnetic moment in these materials could not be induced by so-called insufficiently diluted
magnetic impurities, some physicists remained skeptical~\cite{Mak06}. 
We believe that one of the reasons is no convincing theory to adequately 
explain the magnetism induced by $sp$-electrons.

It is well known that magnetism in materials with TM or REM resulted from large exchange
energy of $d$ or $f$-electrons~\cite{Vol10}. However, $sp$-electrons are commonly considered to have
small exchange energy~\cite{Vol10}. As such, because the $sp$-electrons in these materials
behave in a localized fashion, just like in an isolated atom, it was suggested that Hund's rule could
be used to understand the magnetism~\cite{Vol10}. If the magnetism in materials
without TM or REM was observed by experiments, the magnetism may be understood by Hund's
rule, because atom-like-localized electrons can really exist in real materials. However, if
magnetism emerges from a one-body band structure calculation, we need to be careful with the
interpretation of the origin of the magnetism. Within the framework of band theory, the eigenstates
are Bloch modes that are extended. As such, the magnetism arising from band theory should not
be attributed to localized electrons per se, because localized electrons do not exist in calculations
based on the single electron approximation and Bloch theorem. Even if the dispersion of a band
looks flat, the electrons on this band are still extended, moving everywhere in crystal. This may be
why skeptical scientists~\cite{Mak06} are not convinced 
of the magnetization of materials without TM or REM. 

With $g$-C$_4$N$_3$ as an example, we revealed for the first time that the magnetism in nominally non-magnetic materials 
(including only $sp$-electrons) can be traced to partially filling antibonding states and the
antisymmetric fashion of their wavefunctions.
Since the antibonding wavefunctions are spatially antisymmetric, 
the spin wavefunctions should be symmetric according to the exchange antisymmetric principle of quantum mechanics. 
This is a conclusion distinctly different from the previous investigations about magnetism in non-magnetic 
materials as the mentioned above. 

The results were obtained through first principles calculations
performed on the framework of density functional theory with the projector augmented
plane-wave potential~\cite{PAW} as implemented in the VASP package~\cite{VASP}. The
exchange-correlation term was described by local density approximation (LDA)~\cite{LDA}
which treated the atomic positions well in the carbon-based structures~\cite{Wang09}. The wave
functions were expanded in a plane-wave basis with an energy cutoff of 600~eV. The $k$ points in
surface Brillouin zone (BZ) were sampled on a $\Gamma$-centered $11\times 11$ mesh. The
repeated slab has a vacuum layer of 20~\AA. All atoms were relaxed until the Hellmann-Feynman
forces on the atoms were less than 0.001~eV/\AA. We performed the maximally localized
Wannier function (MLWF) process implemented in Wannier90 package~\cite{MLWF} 
to analyze the orbital components of the concerned bands.

Following its successful synthesis~\cite{Lee10}, $g$-C$_4$N$_3$ has attracted considerable
attention since Du {\it et al}.~\cite{Du12} employed first
principles calculations to identify how $g$-C$_4$N$_3$ exhibited a ferromagnetic ground state
and intrinsic half metallicity, giving it potential applications in spintronics. The 1~$\mu_B$
magnetic moment of $g$-C$_4$N$_3$ was attributed to the contribution of N's
$p$-electrons~\cite{Du12}. 
The band structures of
$g$-C$_4$N$_3$ given by Du {\it et al} indicated that the bands near the Fermi level were
largely dispersive~\cite{Du12} and, as such the magnetism holding states did not behave like an
isolated atom. 
Therefore, $g$-C$_4$N$_3$ is a typical example that can be employed to
investigate the magnetism induced by $sp$-electrons.
Clarifying the physics behind the spin polarization of $sp$-electrons is important for potential
applications of these forms of materials in spintronics devices\cite{Khe14,Pes12,Han14}, because this type of
spin polarization is compatible with many current matured technologies, which mainly rely on a
main group semiconductor.

\begin{figure}[bt]
\centerline{\includegraphics[scale=1.00,angle=0]{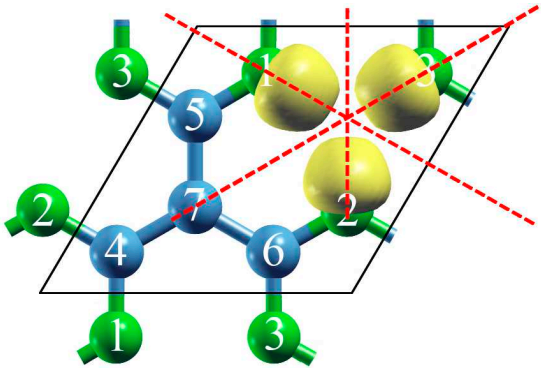}}
\caption{(Color online)
Atomic configuration of $g$-C$_4$N$_3$. Blue and green balls represent C and N atoms
respectively, yellow lobes show the dangling bonds on the N atoms, 
black lines represent the boundary of the unit cell, and the three red dashed lines indicate
the axes of the mirror-plane.
}
\label{config}
\end{figure}

The atomic structure of $g$-C$_4$N$_3$ was optimized and shown in Fig.~\ref{config}. As can be seen
$g$-C$_4$N$_3$ is arranged in a $2\times 2$ honeycomb structure with a vacancy. The lattice
constant of the structure is 4.79~\AA\ and the bond lengths of C-C and C-N are 1.42 and 1.34~\AA\
respectively, in agreement with those of the experiments~\cite{Lee10} and the first principles
calculations~\cite{Du12}. 
No Jahn-Teller distortion (JTD) was found in $g$-C$_4$N$_3$. 
It can be expected because N has one electron more than that of C atom. 
The small discrepancy of the lattice constant we obtained with LDA
compared to the previous calculations with GGA~\cite{Du12} could be attributed to the
difference between LDA and GGA interaction. It has been found LDA could effectively describe
the interaction involved with C atoms~\cite{Wang09}, as such LDA was adopted in this work. We
also examined the half metallicity and other properties by using the exchange-correlation form of
GGA~\cite{GGA}. The results of this investigation revealed that these properties do not depend
on the form of exchange-correlation.

The largest displacement of the C and N to the two dimensional atom plane was smaller than
$0.5*10^{-4}$~\AA. Therefore, all four C and three N atoms should be well $sp^2$ hybridized
and form $\sigma$ bonds between C and N atoms, as well as between C and C atoms. Interactions
between $p_z$ orbitals of all C and N atoms are complex since $\pi$-electrons are no longer pairwise
coupled due to a vacancy in the $2\times 2$ honeycomb structure. This complex and its influence
on magnetism will be discussed in more detail below. With the exception of the $p_z$ orbitals, the
three N atoms connect with three neighbor C atoms in $\sigma$ bonds. The N atoms have dangling
bonds, and the interaction between these plays a critical role in the magnetism in
$g$-C$_4$N$_3$, as discussed below.

\begin{figure}[bt]
\centerline{\includegraphics[scale=0.40,angle=0]{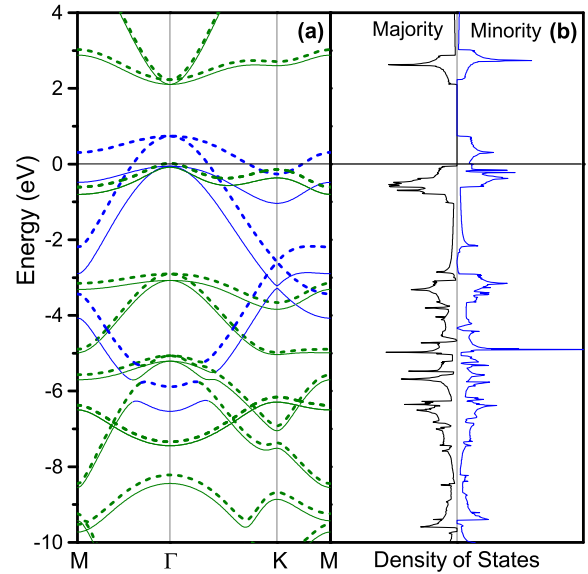}}
\caption{(Color online)
(a) Band structure and (b) density of states of $g$-C$_4$N$_3$. 
Majority and minority bands in (a) are represented by thin solid and thick dashed lines
respectively.  
The Fermi level is set at zero.
}
\label{band-C4N3}
\end{figure}

The stability of $g$-C$_4$N$_3$ was examined by Du {\it et al}. in terms of first principles
molecular dynamics~\cite{Du12}. Thus, we focus on $g$-C$_4$N$_3$'s electronic structures.
The band structures of $g$-C$_4$N$_3$ along the high symmetric axes M-$\Gamma$-K-M are
shown in Fig.~\ref{band-C4N3}. 
Majority and minority bands are represented by thin solid and thick dashed lines respectively.
In agreement with the work of Du {\it et al}.~\cite{Du12},
our calculations revealed that $g$-C$_4$N$_3$ is a semiconductor with $+1e$ hole with a gap of
about 2.16~eV. The highest band of the majority bands just touches the Fermi level at $\Gamma$,
above which a gap of 2~eV for the majority bands exists. The calculated magnetic moment is
1.0~$\mu_B$, indicating its half metallic feature. This is, again, in agreement with the previous
calculations~\cite{Du12}. 

The density of states shown in Fig.~\ref{band-C4N3} (b) indicated that the magnetic moment contribution mainly
originates from the states near the Fermi level, corresponding to the blue bands
in Fig.~\ref{band-C4N3} (a). These are doublet
degenerated at the $\Gamma$ point with character $e$ of group $C_{3V}$. 
Contrary to the Du {\it et al}.'s
calculations of the contribution of N's $p$-electrons~\cite{Du12}, our analysis of both the projected
wavefunction on atomic sites in vasp and the MLWF
fitting indicates that the blue bands can be traced to the $sp^2$ orbitals of the three N atoms.
We note that the conclusion of the $sp^2$ hybridization is reasonable, since all N and C atoms are
on the same plane, therefore, with the exception of the $p_z$ orbitals, the other orbitals of both C
and N atoms should be in the $sp^2$ hybridization. That is, the orbitals $s$, $p_x$ and $p_y$ form a
$sp^2$ hybridization, supposing that the $xy$-plane is the atomic plane of $g$-C$_4$N$_3$.

Clearly, the blue bands represent the interaction between the three dangling bonds of the three N
atoms. 
As a results of our previous investigation into Au(Bi) on
$g$-C$_{14}$N$_3$~\cite{Yu161,Yu162}, which is similar to $g$-C$_4$N$_3$ but in a more
extended network of a $3\times 3$ size graphene, we were aware that the dangling bonds of the three N
atoms interact with each other to form one singlet bonding and two doublet antibonding states.
Through examining the orbital components, we observed a
similar phenomenon in $g$-C$_4$N$_3$. By means of the nearest neighbor tight-binding model
we can set up the Hamiltonian for the interaction between three dangling bonds
\begin{eqnarray*}
\label{eqTB}
\left(
\begin{array}{ccc}
\epsilon_{sp^2}&-T&-T\\
-T&\epsilon_{sp^2}&-T\\
-T&-T&\epsilon_{sp^2}\\
\end{array}
\right),
\end{eqnarray*}
here $\epsilon_{sp^2}$ and $T$ represent, respectively, on-site energy of the dangling bonds and the
interaction between them. Thus, we obtained the singlet bonding energy,
$E_{\rm bonding}=\epsilon_{sp^2}-2T$
and the doublet antibonding energy,
$E_{\rm antibonding}=\epsilon_{sp^2}+2T$.
The singlet bonding state and the doublet antibonding states correspond to the three blue bands at $\Gamma$-point,
as shown in Fig.~\ref{band-C4N3} (a).
The interaction between three dangling bonds would form one bonding and two antibonding states, 
which are critical for the conclusion (see below) about magnetism in materials with only $sp$-electrons 
and do not depend on the form of exchange-correlation.

\begin{figure}[bt]
\centerline{\includegraphics[scale=0.20,angle=0]{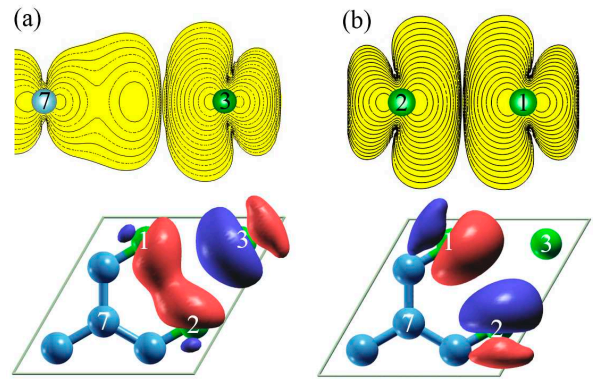}}
\caption{(Color online)
Illustration of antisymmetric combinations of three N's dangling bonds for the antibonding states 
represented by the blue solid lines near the Fermi level in Fig.~\ref{band-C4N3} (a). (a) and (b) for 
the mirror-symmetric and mirror-antisymmetric orbital combination respectively for one of 
the three mirror-planes shown in Fig.~\ref{config}. 
Top: charge density on a plane cutting vertically through the connection between the labeled atoms. 
Bottom: schematic representation of the mirror-symmetric and mirror antisymmetric combinations 
for one of the three mirror-planes. The red and blue lobes indicate the orbitals with  
opposite sign. 
}
\label{charge-density}
\end{figure}

As is the case with $g$-C$_{14}$N$_3$~\cite{Yu161,Yu162}, the two higher-lying blue bands, which are
doublet degenerate at $\Gamma$ and separated apart from the $\Gamma$ point, are the
antibonding states of the three N's dangling bonds with character $e$ of group $C_{3V}$ in a
mirror-symmetry combination (MSC) and mirror-antisymmetry combination (MAC). 
In order to examine the nature of antisymmetric combination, 
the charge densities for the two antibonding states at $\Gamma$ 
were calculated and plotted in Fig.~\ref{charge-density}. Fig.~\ref{charge-density} (a) and (b) 
reveal MSC and MAC features, respectively, for one of three mirror-planes. 
We found that, 
regardless of the value to which the contour was adjusted, 
the contour lines could not continually extend from atoms 7 to 3 for MSC 
as well as from atoms 2 to 1 for MAC, indicating nodes existed and 
the dangling bond orbitals of N atoms 1, 2 and 3 must antisymmetrically 
consist to form the state. Our MLWF analysis also confirmed the orbitals 
with opposite sign, as schematically shown by red and blue lobes 
in the bottom panels of Fig.~\ref{charge-density}.
Notably, the
two $e$-bands (antibonding states) appear near the Fermi level. Furthermore, its majority bands are
fully occupied, while its minority ones are partially occupied, indicating spin polarization.

If the main magnetic moment originates from the partially occupied antibonding states, it is
natural to question: in which form do electrons favor to occupy the antibonding states, spin
parallel or spin antiparallel? It is well known that the wavefunction of the bonding state consists of
the wavefunctions of the bonding atoms in a symmetric form, while that of the antibonding state
consists of the wavefunctions in an antisymmetric form. That is, the coordinates of an antibonding
wavefunction are naturally antisymmetric. According to quantum mechanics, the wavefunctions of
electrons should be antisymmetric with respect to the interchange of electron coordinates and
spins. 
If the coordinates
of the wavefunction are antisymmetric, the spins should be symmetric, and vice versa. The
coordinates of an antibonding wavefunction are naturally antisymmetric. As such, their spins
should be parallel. This is the physics behind the spin polarization of $g$-C$_4$N$_3$.

It was reported that in graphene with vacancies, two of the three atoms around the vacancy 
form a $sp^{2}\sigma$ bond due to JTD, leaving the apical atom with a dangling bond contributing 
a magnetic moment of approximately 1.0$\mu_{\rm B}$~\cite{Leh04,Ma04}.
However, we emphasize that $g$-C$_4$N$_3$ is different, although the atomic structure of $g$-C$_4$N$_3$ 
is similar (three C replaced by three N) to that of graphene with vacancy. 
The $g$-C$_4$N$_3$ has no JTD and the concerned bands have dispersions as large as 2.5eV, 
as shown in Fig.~\ref{band-C4N3}, meaning their dislocalized nature.

In summary, through the example of $g$-C$_4$N$_3$, this study revealed that the 
origin of the magnetism in $g$-C$_4$N$_3$ can be traced to the electron occupation of the 
antibonding states in spin polarization. 
Therefore, the present work can end a long history of dispute in this field.
Since the antibonding wavefunctions are spatially 
antisymmetric, the spins should be symmetric according to the principle of the exchange 
antisymmetric principle of quantum mechanics. These results allow us to conclude that the 
magnetism of otherwise non-magnetic materials is possible, at least in materials that have a
vacancy structure. Here a crucial point is that the antibonding states should appear near the Fermi 
level and should be partially occupied. 

This work was supported by NFSC (No.61274097) and NBRPC (No. 2015CB921401).

\end{document}